# Probabilistically Bounded Staleness for Practical Partial Quorums

Peter Bailis, Shivaram Venkataraman, Michael J. Franklin, Joseph M. Hellerstein, Ion Stoica
University of California, Berkeley
{pbailis, shivaram, franklin, hellerstein, istoica}@cs.berkeley.edu

*All good ideas arrive by chance.*—Max Ernst

## ABSTRACT

Data store replication results in a fundamental trade-off between operation latency and data consistency. In this paper, we examine this trade-off in the context of quorum-replicated data stores. Under partial, or non-strict quorum replication, a data store waits for responses from a subset of replicas before answering a query, without guaranteeing that read and write replica sets intersect. As deployed in practice, these configurations provide only basic eventual consistency guarantees, with no limit to the recency of data returned. However, anecdotally, partial quorums are often "good enough" for practitioners given their latency benefits. In this work, we explain why partial quorums are regularly acceptable in practice, analyzing both the staleness of data they return and the latency benefits they offer. We introduce Probabilistically Bounded Staleness (PBS) consistency, which provides expected bounds on staleness with respect to both versions and wall clock time. We derive a closed-form solution for versioned staleness as well as model real-time staleness for representative Dynamo-style systems under internet-scale production workloads. Using PBS, we measure the latency-consistency trade-off for partial quorum systems. We quantitatively demonstrate how eventually consistent systems frequently return consistent data within tens of milliseconds while offering significant latency benefits.

## 1. INTRODUCTION

Modern distributed data stores need to be scalable, highly available, and fast. These systems typically replicate data across different machines and often across datacenters for two reasons: first, to provide high availability when components fail and, second, to provide improved performance by serving requests from multiple replicas. In order to provide predictably low read and write latency, systems often eschew protocols guaranteeing consistency of reads[1] and instead opt for eventually consistent protocols [4, 6, 20, 23, 38, 39, 55]. However, eventually consistent systems make no guarantees on the staleness (recency in terms of versions written) of data items returned except that the system will "eventually" return the most recent version in the absence of new writes [61].

This latency-consistency trade-off inherent in distributed data stores has significant consequences for application design [6]. Low latency is critical for a large class of applications [56]. For example, at Amazon, 100 ms of additional latency resulted in a 1% drop in sales [44], while 500 ms of additional latency in Google's search resulted in a corresponding 20% decrease in traffic [45]. At scale, increased latencies correspond to large amounts of lost revenue, but lowering latency has a consistency cost: contacting fewer replicas for each request typically weakens the guarantees on returned data. Programs can often tolerate weak consistency by employing careful design patterns such as compensation (e.g., memories, guesses, and apologies) [33] and by using associative and commutative operations (e.g., timelines, logs, and notifications) [12]. However, potentially *unbounded* staleness (as in eventual consistency) poses significant challenges and is undesirable in practice.

### 1.1 Practical Partial Quorums

In this work, we examine the latency-consistency trade-off in the context of quorum-replicated data stores. Quorum systems ensure strong consistency across reads and writes to replicas by ensuring that read and write replica sets overlap. However, employing *partial* (or non-strict) quorums can lower latency by requiring fewer replicas to respond. With partial quorums, sets of replicas written to and read from need not overlap: given $N$ replicas and read and write quorum sizes $R$ and $W$, partial quorums imply $R+W \leq N$.

Quorum-replicated data stores such as Dynamo [20] and its open source descendants Apache Cassandra [41], Basho Riak [3], and Project Voldemort [24] offer a choice between two modes of operation: strict quorums with strong consistency or partial quorums with eventual consistency. Despite eventual consistency's weak guarantees, operators frequently employ partial quorums [1, 4, 23, 38, 55, 64]—a controversial decision [32, 46, 57, 58]. Given their performance benefits, which are especially important as latencies grow [6, 23, 32, 33], partial quorums are often considered acceptable. The proliferation of partial quorum deployments suggests that applications can often tolerate occasional cases of staleness and that data tends to be "fresh enough" in most cases.

While common practice suggests that eventual consistency is often a viable solution for operators, to date, this observation has been anecdotal. In this work, we quantify the degree to which eventual consistency is both eventual and consistent and explain why. Under worst-case conditions, eventual consistency results in an unbounded degree of data staleness, but, as we will show, the average case is often different. Eventually consistent data stores cannot promise immediate and perfect consistency but, for varying degrees of certainty, can offer staleness bounds with respect to time ("how eventual") and version history ("how consistent").

---

[1]This distributed replica consistency differs from transactional consistency provided by ACID semantics [50, 58].

There is little prior work describing how to make these consistency and staleness predictions under practical conditions. The current state of the art requires that users make rough guesses or perform online profiling to determine the consistency provided by their data stores [16, 28, 62]. Users have little to no guidance on how to chose an appropriate replication configuration or how to predict the behavior of partial quorums in production environments.

## 1.2 PBS Predictions and Contributions

To predict consistency, we need to know when and why eventually consistent systems return stale data and how to quantify the staleness of the data they return. In this paper, we answer these questions by expanding prior work on *probabilistic quorums* [49, 51] to account for multi-version staleness and message dissemination protocols as used in today's systems. More precisely, we present algorithms and models for predicting the staleness of partial quorums, called Probabilistically Bounded Staleness (PBS). There are two common metrics for measuring staleness in the literature: wall clock time [28, 65, 66] and versions [28, 40, 67]. PBS describes both measures, providing the probability of reading a write $t$ seconds after it returns ($t$-visibility, or "how eventual is eventual consistency?"), of reading one of the last $k$ versions of a data item ($k$-staleness, or "how consistent is eventual consistency?"), and of experiencing a combination of the two ($\langle k, t \rangle$-staleness). PBS does not propose new mechanisms to enforce deterministic staleness bounds [40, 54, 65, 66, 67]; instead, our goal is to provide a lens for analyzing, improving, and predicting the behavior of *existing*, widely deployed systems.

We provide closed-form solutions for PBS $k$-staleness and use Monte Carlo methods to explore the trade-off between latency and $t$-visibility. We present a detailed study of Dynamo-style PBS $t$-visibility using production latency distributions. We show how long-tailed one-way write latency distributions affect the time required for a high probability of consistent reads. For example, in one production environment, switching from spinning disks to solid-state drives dramatically improved staleness (e.g., 1.85ms versus 45.5ms wait time for a 99.9% probability of consistent reads) due to decreased write latency mean and variance. We also make quantitative observations of the latency-consistency trade-offs offered by partial quorums. For example, in another production environment, we observe an 81.1% combined read and write latency improvement at the 99.9th percentile (230 to 43.3ms) for a 202ms window of inconsistency (99.9% probability consistent reads). This analysis demonstrates the performance benefits that lead operators to choose eventual consistency.

We make the following contributions in this paper:

- We develop the theory of Probabilistically Bounded Staleness (PBS) for partial quorums. PBS describes the probability of staleness across versions ($k$-staleness) and time ($t$-visibility) as well as the probability of session-based monotonic reads consistency.

- We provide a closed-form analysis of $k$-staleness demonstrating how the probability of receiving data $k$ versions old is exponentially reduced by $k$. As a corollary, $k$-staleness tolerance also exponentially lowers quorum system *load*.

- We describe the *WARS* model for $t$-visibility in Dynamo-style partial quorum systems and show how message reordering leads to staleness. We evaluate the $t$-visibility of Dynamo-style systems using a combination of synthetic and production latency models.

## 2. BACKGROUND

In this section, we provide background on quorum systems both in the theoretical academic literature and in practice. We begin by introducing prior work on traditional and probabilistic quorum systems. We next discuss Dynamo-style quorums, currently the most widely deployed protocol for data stores employing quorum replication. Finally, we survey reports of practitioner usage of partial quorums for three Dynamo-style data stores.

### 2.1 Quorum Foundations: Theory

Systems designers have long proposed quorum systems as a replication strategy for distributed data [26]. Under quorum replication, a data store writes a data item by sending it to a set of replicas, called a write quorum. To serve reads, the data store fetches the data from a possibly different set of replicas, called a read quorum. For reads, the data store compares the set of values returned by the replicas, and, given a total ordering of versions,[2] can return the most recent value (or all values received, if desired). For each operation, the data store chooses read and write quorums from a set of sets of replicas, known as a *quorum system*, with one system per data item. There are many kinds of quorum systems, but one simple configuration is to use read and write quorums of fixed sizes, which we will denote $R$ and $W$, for a set of nodes of size $N$. To reiterate, a quorum replicated data store uses one quorum system per data item. Across data items, quorum systems need not be identical

Informally, a strict quorum system is a quorum system with the property that any two quorums (sets) in the quorum system overlap (have non-empty intersection). This ensures consistency. The minimum sized quorum defines the system's fault tolerance, or availability. A simple example of a strict quorum system is the majority quorum system, in which each quorum is of size $\lceil \frac{N}{2} \rceil$. The theory literature describes alternative quorum system designs providing varying asymptotic properties of capacity, scalability, and fault tolerance, from tree-quorums [8] to grid-quorums [52] and highly available hybrids [9]. Jiménez-Peris et al. provide an overview of traditional, strict quorum systems [37].

Partial quorum systems are natural extensions of strict quorum systems: at least two quorums in a partial quorum system do not overlap. There are two relevant variants of partial quorum systems in the literature: probabilistic quorum systems and k-quorums.

*Probabilistic quorum systems* provide probabilistic guarantees of quorum intersection. By scaling the number of replicas, one can achieve an arbitrarily high probability of consistency [49]. Intuitively, this is a consequence of the Birthday Paradox: as the number of replicas increases, the probability of non-intersection between any two quorums decreases. Probabilistic quorums are typically used to predict the probability of strong consistency but not (multi-version) bounded staleness. Merideth and Reiter provide an overview of these systems [51].

As an example of a probabilistic quorum system, consider $N$ replicas with randomly chosen read and write quorums of sizes $R$ and $W$. We can calculate the probability that the read quorum does not contain the last written version. This probability is the number of quorums of size $R$ composed of nodes that were not written to in the write quorum divided by the number of possible read quorums:

$$p_s = \frac{\binom{N-W}{R}}{\binom{N}{R}} \quad (1)$$

---

[2]We can easily achieve a total ordering using globally synchronized clocks or using a causal ordering provided by mechanisms such as vector clocks [42] with commutative merge functions [46]

777

The probability of inconsistency is high except for large $N$. With $N = 100$, $R = W = 30$, $p_s = 1.88 \times 10^{-6}$ [10]. However, with $N = 3$, $R = W = 1$, $p_s = 0.\overline{6}$. The asymptotics of these systems are excellent—but only asymptotically.

*k-quorum systems* provide *deterministic* guarantees that a partial quorum system will return values that are within $k$ versions of the most recent write [10]. In a single writer scenario, sending each write to $\lceil \frac{N}{k} \rceil$ replicas with round-robin write scheduling ensures that any replica is no more than $k$ versions out-of-date. However, with multiple writers, we lose the global ordering properties that the single-writer was able to control, and the best-known algorithm for the pathological case results in a lower bound of $(2N-1)(k-1) + N$ versions staleness [11].

This prior work makes two important assumptions. First, it typically models quorum sizes as fixed, where the set of nodes with a version does not grow over time. Prior work examined "dynamic systems", considering quorum membership churn [7], network-aware quorum placement [25, 29], and network partitions [34] but not write propagation. Second, it frequently assumes Byzantine failure. We revisit these assumptions in the next section.

## 2.2 Quorum Foundations: Practice

In practice, many distributed data management systems use quorums as a replication mechanism. Amazon's Dynamo [20] is the progenitor of a class of eventually consistent key-value stores that includes Apache Cassandra [41], Basho Riak [3], and LinkedIn's Project Voldemort [24]. All use the same variant of quorum-style replication and we are not aware of any widely adopted data store using a vastly different quorum replication protocol. However, with some work, we believe that other styles of replication can adopt our methodology. We describe key-value stores here, but any replicated data store can use quorums, including full RDBMS systems.

Dynamo-style quorum systems employ one quorum system per key, typically maintaining the mapping of keys to quorum systems using a consistent-hashing scheme or a centralized membership protocol. Each node stores multiple keys. As shown in Figure 1, clients send read and write requests to a node in the system cluster, which forwards the request to *all* nodes assigned to that key as replicas. This coordinating node considers an operation complete when it has received responses from a pre-determined number of replicas (typically set per-operation). Accordingly, without message loss, all replicas eventually receive all writes. This means that the write and read quorums chosen for a request depend on which nodes respond to the request first. Dynamo denotes the replication factor of a key as $N$, the number of replica responses required for a successful read as $R$, and the number of replica acknowledgments required for a successful write as $W$. Under normal operation, Dynamo-style systems guarantee consistency when $R+W>N$. Setting $W>\lceil N/2 \rceil$ ensures consistency in the presence of concurrent writes.

There are significant differences between quorum theory and data systems used in practice. First, replication factors for data stores are low, typically between one and three [4, 23, 30]. Second, (in the absence of failure), in Dynamo-style partial quorums, the write quorum size increases even after the operation returns, growing via anti-entropy [21]. Coordinators send all requests to all replicas but consider only the first $R$ ($W$) responses. As a matter of nomenclature (and to disambiguate against "dynamic" quorum membership protocols), we will refer to these systems as *expanding partial quorum systems*. (We discuss additional anti-entropy in Section 4.2.) Third, as in much of the applied literature, practitioners focus on fail-stop instead of Byzantine failure modes [17]. Following standard practice, we do not consider Byzantine failure.

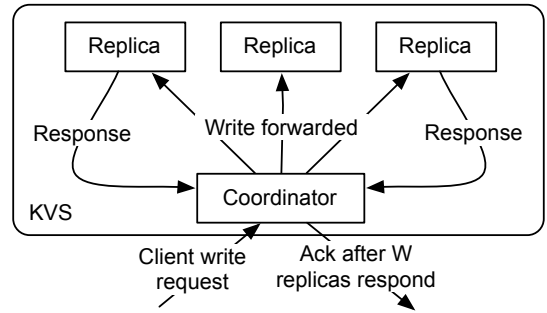

**Figure 1: Diagram of control flow for client write to Dynamo-style quorum ($N = 3$, $W = 2$). A coordinator node handles the client write and sends it to all $N$ replicas. The write call returns after the coordinator receives $W$ acknowledgments.**

## 2.3 Typical Quorum Configurations

For improved latency, operators often set $R+W \leq N$. Here, we survey quorum configurations according to practitioner accounts. Operators frequently use partial quorum configurations, citing performance benefits and high availability. Most of these accounts did not discuss the possibility or occurrence of staleness resulting from partial quorum configurations.

Cassandra defaults to $N=3$, $R=W=1$ [4]. The Apache Cassandra 1.0 documentation claims that "a majority of users do writes at consistency level [$W=1$]", while the Cassandra Query Language defaults to $R=W=1$ as well [1]. Production Cassandra users report using $R=W=1$ in the "general case" because it provides "maximum performance" [64], which appears to be a commonly held belief [38, 55]. Cassandra has a "minor" patch [2] for session guarantees [60] that is not currently used [22]; according to our discussions with developers, this is due to lack of interest.

Riak defaults to $N=3$, $R=W=2$ [14, 15]. Users suggest using $R=W=1$, $N=2$ for "low value" data (and strict quorum variants for "web," "mission critical," and "financial" data) [39, 47].

Finally, Voldemort does not provide sample configurations, but Voldemort's authors (and operators) at LinkedIn [23] often choose $N=c$, $R=W=\lceil c/2 \rceil$ for odd $c$. For applications requiring "very low latency and high availability," LinkedIn deploys Voldemort with $N=3$, $R=W=1$. For other applications, LinkedIn deployments Voldemort with $N=2$, $R=W=1$, providing "some consistency," particularly when three-way replication is not required. Unlike Dynamo, Voldemort sends read requests to $R$ of $N$ replicas (not $N$ of $N$) [24]; this decreases load per replica and network traffic at the expense of read latency and potential availability. Provided staleness probabilities are independent across requests, this does not affect staleness: even when sending reads to $N$ replicas, coordinators only wait for $R$ responses.

## 3. PROBABILISTICALLY BOUNDED STALENESS

In this section, we introduce Probabilistically Bounded Staleness, which describes the consistency provided by existing eventually consistent data stores. We present PBS $k$-staleness, which probabilistically bounds the staleness of versions returned by read quorums, PBS $t$-visibility, which probabilistically bounds the time before a committed version appears to readers, and PBS $\langle k,t \rangle$-staleness, a combination of the two prior models.

We introduce $k$-staleness first because it is self-contained, with a simple closed-form solution. In comparison, $t$-visibility is more difficult, involving additional variables. Accordingly, this section proceeds in order of increasing difficulty, and the remainder of the paper addresses the complexities of $t$-visibility.

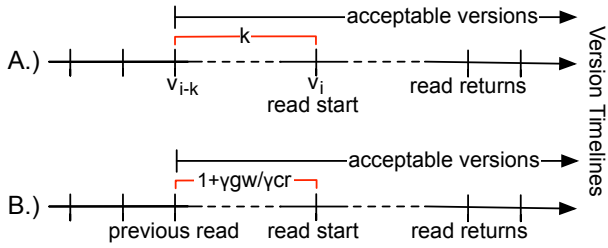

**Figure 2:** Versions returnable by read operations under PBS $k$-staleness (A) and PBS monotonic reads (B). In $k$-staleness, the read operation will return a version no later than $k$ versions older than the last committed value when it started. In monotonic reads consistency, acceptable staleness depends on the number of versions committed since the client's last read.

Practical concerns guide the following theoretical contributions. We begin by considering a model without quorum expansion or other anti-entropy. For the purposes of a running example, as in Equation 1, we assume that $W$ ($R$) of $N$ replicas are randomly selected for each write (read) operation. Similarly, we consider fixed $W$, $R$ and $N$ across multiple operations. Next, we expand our model to consider write propagation and time-varying $W$ sizes in expanding partial quorums. In this section, we discuss anti-entropy in general, however we model Dynamo-style quorums in Section 4. We discuss further refinements to these assumptions in Section 6.

### 3.1 PBS $k$-staleness

Probabilistic quorums allow us to determine the probability of returning the most recent value written to the database, but do not describe what happens when the most recent value is not returned. Here, we determine the probability of returning a value within a bounded number of versions. In the following formulation, we consider traditional, non-expanding write quorums (no anti-entropy):

*Definition 1.* A quorum system obeys *PBS k-staleness consistency* if, with probability $1 - p_{sk}$, at least one value in any read quorum has been committed within $k$ versions of the latest committed version when the read begins.

Reads may return versions whose writes that are not yet committed (in-flight) (see Figure 2A). The $k$-quorum literature defines these as $k$-regular semantics [10].

The probability of returning a version of a key within the last $k$ versions committed is equivalent to intersecting one of $k$ independent write quorums. Given the probability of a single quorum non-intersection $p$, the probability of non-intersection with one of the last $k$ independent quorums is $p^k$. In our running example, the probability of non-intersection is Equation 1 exponentiated by $k$:

$$p_{sk} = \left( \frac{\binom{N-W}{R}}{\binom{N}{R}} \right)^k \quad (2)$$

When $N=3$, $R=W=1$, this means that the probability of returning a version within 2 versions is $0.\overline{5}$, within 3 versions, $0.\overline{703}$, 5 versions, $> 0.868$, and 10 versions, $> 0.98$. When $N=3$, $R=1$, $W=2$ (or, equivalently, $R=2$, $W=1$), these probabilities increase: $k=1 \to 0.\overline{6}$, $k=2 \to 0.\overline{8}$, and $k=5 \to > 0.995$.

This closed form solution holds for quorums that do not change size over time. For expanding partial quorum systems, this solution is an upper bound on the probability of staleness.

### 3.2 PBS Monotonic Reads

PBS $k$-staleness can predict whether a client will ever read older data than it has previously read, a well-known session guarantee called *monotonic reads* consistency [60]. This is particularly useful when clients do not need to see the most recent version of a data item but still require a notion of "forward progress" through versions, as in timelines or streaming change logs.

*Definition 2.* A quorum system obeys *PBS monotonic reads consistency* if, with probability at least $1 - p_{sMR}$, at least one value in any read quorum returned to a client is the same version or a newer version than the last version that the client previously read.

To guarantee that a client sees monotonically increasing versions, it can continue to contact the same replica [61] (provided the "sticky" replica does not fail). However, this is insufficient for strict monotonic reads (where the client reads strictly newer data if it exists in the system). We can adapt Definition 2 to accommodate strict monotonic reads by requiring that the data store returns a more recent data version if it exists.

PBS monotonic reads consistency is a special case of PBS $k$-staleness (see Figure 2B), where $k$ is determined by a client's rate of reads from a data item ($\gamma_{cr}$) and the global, system-wide rate of writes to the same data item ($\gamma_{gw}$). If we know these rates, the number of versions written between client reads is $\frac{\gamma_{gw}}{\gamma_{cr}}$, as shown in Figure 2B. We can calculate the probability of probabilistic monotonic reads as a special case of $k$-staleness where $k = 1 + \frac{\gamma_{gw}}{\gamma_{cr}}$. Again extending our running example, from Equation 2:

$$p_{sMR} = \left( \frac{\binom{N-W}{R}}{\binom{N}{R}} \right)^{1+\gamma_{gw}/\gamma_{cr}} \quad (3)$$

For strict monotonic reads, where we cannot read the version we have previously read (assuming there are newer versions in the database), we exponentiate with $k = \frac{\gamma_{gw}}{\gamma_{cr}}$.

In practice, we may not know these exact rates, but, by measuring their distribution, we can calculate an expected value. By performing appropriate admission control, operators can control these rates to achieve monotonic reads consistency with high probability.

### 3.3 Load Improvements

Theory literature defines the *load* of a quorum system as a metric for the frequency of accessing the busiest quorum member [52, Definition 3.2]. Intuitively, the busiest quorum member limits the number of requests that a given quorum system can sustain, called its *capacity* [52, Corollary 3.9].

Prior work determined that probabilistic quorum systems did not offer significant benefits to load (providing a constant factor improvement compared to strict quorum systems) [49]. Here, we show that quorums tolerating PBS $k$-staleness have asymptotically lower load than traditional probabilistic quorum systems (and, transitively, than strict quorum systems).

The probabilistic quorum literature defines an $\varepsilon$-intersecting quorum system as a quorum system that provides a $1 - \varepsilon$ probability of returning consistent data [49, Definition 3.1]. A $\varepsilon$-intersecting quorum system has load of at least $\frac{1-\sqrt{\varepsilon}}{\sqrt{N}}$ [49, Corollary 3.12].

In considering $k$ versions of staleness, we consider the intersection of $k$ $\varepsilon$-intersecting quorum systems. For a given probability $p$ of inconsistency, if we are willing to tolerate $k$ versions of staleness, we need only require that $\varepsilon = \sqrt[k]{p}$. This implies that our PBS $k$-staleness system construction has load of at least $\frac{(1-p)^{\frac{1}{2k}}}{\sqrt{N}}$, an

779

improved lower bound compared to traditional probabilistic quorum systems. PBS monotonic reads consistency results in a lower bound on load of $\frac{(1-p)^{\frac{1}{2C}}}{\sqrt{N}}$, where $C = 1 + \frac{\gamma_{gw}}{\gamma_{cr}}$.

These results are intuitive: if we are willing to tolerate multiple versions of staleness, we need to contact fewer replicas. Staleness tolerance lowers the load of a quorum system, subsequently increasing its capacity.

### 3.4 PBS $t$-visibility

Until now, we have considered only quorums that do not grow over time. However, as we discussed in Section 2.2, real-world quorum systems expand by asynchronously propagating writes to quorum system members over time. This process is commonly known as anti-entropy [21]. For generality, in this section, we will discuss generic anti-entropy. However, we explicitly model the Dynamo-style anti-entropy mechanisms in Section 4.

PBS $t$-visibility models the probability of inconsistency for expanding quorums. $t$-visibility is the probability that a read operation, starting $t$ seconds after a write commits, will observe the latest value of a data item. This $t$ captures the expected length of the "window of inconsistency." Recall that we consider in-flight writes—which are more recent than the last committed version—as non-stale.

*Definition 3.* A quorum system obeys *PBS t-visibility consistency* if, with probability $1 - p_{st}$, any read quorum started at least $t$ units of time after a write commits returns at least one value that is at least as recent as that write.

Overwriting data items effectively resets $t$-visibility; the time between writes bounds $t$-visibility. If we space two writes to a key $m$ milliseconds apart, then the $t$-visibility of the first write for $t > m$ milliseconds is undefined; after $m$ milliseconds, there will be a newer version.

We denote the cumulative density function describing the number of replicas $\mathcal{W}_r$ that have received a particular version $v$ exactly $t$ seconds after $v$ commits as $P_w(\mathcal{W}_r, t)$.

By definition, for expanding quorums, $\forall c \in [0, W], P_w(c, 0) = 1$; at commit time, $W$ replicas will have received the value with certainty. We can model the probability of PBS $t$-visibility for given $t$ by summing the conditional probabilities of each possible $\mathcal{W}_r$:

$$p_{st} = \frac{\binom{N-W}{N}}{\binom{N}{R}} + \sum_{c \in (W,N)} \frac{\binom{N-c}{N}}{\binom{N}{R}} \cdot [P_w(c+1, t) - P_w(c, t)] \quad (4)$$

However, the above equation assumes reads occur instantaneously and writes commit immediately after $W$ replicas have the version (i.e., there is no delay acknowledging the write to the coordinating node). In the real world, coordinators wait for write acknowledgments and read requests take time to arrive at remote replicas, increasing $t$. Accordingly, Equation 4 is a conservative upper bound on $p_{st}$.

In practice, $P_w$ depends on the anti-entropy mechanisms in use and the expected latency of operations but we can approximate it (Section 4) or measure it online. For this reason, the load of a PBS $t$-visible quorum system depends on write propagation and is difficult to analytically determine for general-purpose expanding quorums. Additionally, one can model both transient and permanent failures by increasing the tail probabilities of $P_w$ (Section 6).

### 3.5 PBS $\langle k, t \rangle$-staleness

We can combine the previous models to combine both versioned and real-time staleness metrics to determine the probability that a read will return a value no older than $k$ versions stale if the last write committed at least $t$ seconds ago:

*Definition 4.* A quorum system obeys *PBS $\langle k, t \rangle$-staleness consistency* if, with probability $1 - p_{skt}$, at least one value in any read quorum will be within $k$ versions of the latest committed version when the read begins, provided the read begins $t$ units of time after the previous $k$ versions commit.

The definition of $p_{skt}$ follows from the prior definitions:

$$p_{skt} = \left(\frac{\binom{N-W}{R}}{\binom{N}{R}}\right) + \sum_{c \in [W,N)} \frac{\binom{N-c}{R}}{\binom{N}{R}} \cdot [P_w(c+1, t) - P_w(c, t)]^k \quad (5)$$

In this equation, in addition to (again) assuming instantaneous reads, we also assume the pathological case where the last $k$ writes all occurred at the same time. If we can determine the time since commit for the last $k$ writes, we can improve this bound by considering each quorum's $p_{skt}$ separately (individual $t$). However, predicting (and enforcing) write arrival rates is challenging and may introduce inaccuracy, so this equation is a conservative upper bound on $p_{skt}$.

Note that PBS $\langle k, t \rangle$-staleness consistency encapsulates the prior definitions of consistency. Probabilistic $k$-quorum consistency is simply PBS $\langle k, 0 \rangle$-staleness consistency, PBS monotonic reads consistency is $\langle 1 + \frac{\gamma_{gw}}{\gamma_{cr}}, 0 \rangle$-staleness consistency, and PBS $t$-visibility is $\langle 1, t \rangle$-staleness consistency.

In practice, we believe it is easier to reason about staleness of versions or staleness of time but not both together. Accordingly, having derived a closed-form model for $k$-staleness, in the remainder of this paper, we focus mainly on deriving more specific models for $t$-visibility. A conservative rule-of-thumb going forward is to exponentiate the probability of inconsistency in $t$-visibility by $k$ when up to $k$ versions of staleness are tolerable.

## 4. DYNAMO-STYLE $T$-VISIBILITY

We have a closed-form model for $k$-staleness, but $t$-visibility is dependent on both the quorum replication algorithm and the anti-entropy processes employed by a given system. In this section, we discuss PBS $t$-visibility in the context of Dynamo-style data stores and describe how to asynchronously detect staleness.

### 4.1 Inconsistency in Dynamo: *WARS* Model

Dynamo-style quorum systems are inconsistent as a result of read and write message reordering, a product of message delays. To illustrate this phenomenon, we introduce a model of message latency in Dynamo operation that, for convenience, we call *WARS*.

In Figure 3, we illustrate *WARS* using a space-time diagram for messages between a coordinator and a single replica for a write followed by a read $t$ seconds after the write commits. This $t$ corresponds to the $t$ in PBS $t$-visibility. In brief, reads are stale when all of the first $R$ responses to the read request arrived at their replicas before the last (committed) write request.

For a write, the coordinator sends $N$ messages, one to each replica. The message from the coordinator to replica containing the write is delayed by a value drawn from distribution `W`. The coordinator waits for $W$ responses from the replicas before it can consider the version committed. Each response acknowledging the write is delayed by a value drawn from the distribution `A`.

For a read, the coordinator (possibly different than the write coordinator) sends $N$ messages, one to each replica. The message

780

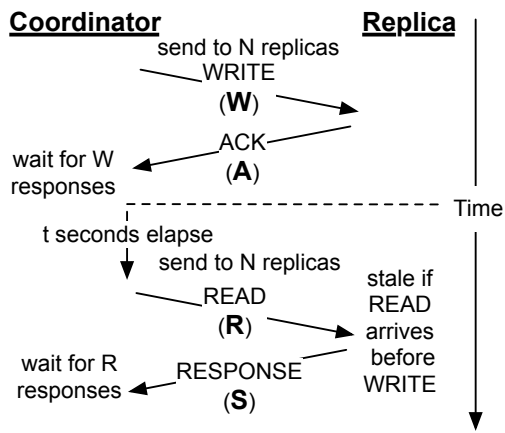

**Figure 3: The *WARS* model for in Dynamo describes the message latencies between a coordinator and a single replica for a write followed by a read $t$ seconds after commit. In an $N$ replica system, this messaging occurs $N$ times.**

from coordinator to replica containing the read request is delayed by a value drawn from distribution R. The coordinator waits for $R$ responses from the replicas before returning the most recent value it receives. The read response from each replica is delayed by a value drawn from the distribution S.

The read coordinator will return stale data if the first $R$ responses received reached their replicas before the replicas received the latest version (delayed by W). When $R+W>N$, this is impossible. However, under partial quorums, the frequency of this occurrence depends on the latency distributions. If we denote the commit time (when the coordinator has received $W$ acknowledgments) as $w_t$, a single replica's response is stale if $r' + w_t + t < w'$ for $r'$ drawn from R and $w'$ drawn from W. Writes have time to propagate to additional replicas both while the coordinator waits for all required acknowledgments (A) and as replicas wait for read requests (R). Read responses are further delayed in transit (S) back to the read coordinator, inducing further possibility of reordering. Qualitatively, longer write tails (W) and faster reads increase the chance of staleness due to reordering.

*WARS* considers the effect of message sending, delays, and reception, but this represents a daunting analytical formulation. The commit time is an order statistic of $W$ and $N$ dependent on both W and A. Furthermore, the probability that the $i$th returned read message observes reordering is another order statistic of $R$ and $N$ dependent on W,A,R, and S. Moreover, across responses, the probabilities are dependent. These dependencies make calculating the probability of staleness rather difficult. Dynamo is straightforward to reason about and program but is difficult to analyze in a simple closed form. As we discuss in Section 5.1, we instead explore *WARS* using Monte Carlo methods, which are straightforward to understand and implement.

### 4.2 *WARS* Scope

**Proxying operations.** Depending on which coordinator a client contacts, coordinators may serve reads and writes locally. In this case, subject to local query processing delays, a read or write to $R$ or $W$ nodes behaves like a read or write to $R-1$ or $W-1$ nodes. Although we do not do so, one could adopt *WARS* to handle local reads and writes. The decision to proxy requests (and, if not, which replicas serve which requests) is data store and deployment-specific. Dynamo forwards write requests to a designated coordinator solely for the purpose of establishing a version ordering [20, Section 6.4] (easily achievable through other mechanisms [36]). Dynamo's authors observed a latency improvement by proxying all operations and having clients act as coordinators—Voldemort adopts this architecture [59].

**Client-side delays.** End-users will likely incur additional time between their reads and writes due to latency required to contact the service. Individuals making requests to web services through their browsers will likely space sequential requests by tens or hundreds of milliseconds due to client-to-server latency. Although we do not consider this delay here, it is important to remember for practical scenarios because delays between reads and writes ($t$) may be large.

**Additional anti-entropy.** As we discussed in Section 2.2, anti-entropy decreases the probability of staleness by propagating writes between replicas. Dynamo-style systems also support additional anti-entropy processes [50]. *Read repair* is a commonly used process: when a read coordinator receives multiple versions of a data item from different replicas in response to a read request, it will attempt to (asynchronously) update the out-of-date replicas with the most recent version [20, Section 5]. Read repair acts like an additional write for every read, except old values are re-written. Additionally, Dynamo used Merkle trees to summarize and exchange data contents between replicas [20, Section 4.7]. However, not all Dynamo-style data stores actively employ similar gossip-based anti-entropy. For example, Cassandra uses Merkle tree anti-entropy only when manually requested (e.g., `nodetool repair`), instead relying primarily on quorum expansion and read repair [5].

These processes are rate-dependent: read repair's efficiency depends on the rate of reads, and Merkle tree exchange's efficiency (and, more generally, most anti-entropy efficiency) depends on the rate of exchange. A conservative assumption for read repair and Merkle tree exchange is that they never occur. For example, assuming a particular read repair rate implies a given rate of reads from each key in the system.

In contrast, *WARS* captures expanding quorum behavior independent of read rate and with conservative write rate assumptions. *WARS* considers a single read and a single write. Aside from load considerations, concurrent reads do not affect staleness. If multiple writes overlap (that is, have overlapping periods where they are in-flight but are not committed) the probability of inconsistency decreases. This is because overlapping writes result in an increased chance that a client reads as-yet-uncommitted data. As a result, with *WARS*, data may be fresher than predicted.

### 4.3 Asynchronous Staleness Detection

Even if a system provides a low probability of inconsistency, applications may need notification when data returned is inconsistent or staler than expected. Here, as a side note, we discuss how the Dynamo protocol is naturally equipped for staleness detection. We focus on PBS $t$-visibility in the following discussion but it is easily extended to PBS $k$-staleness and $\langle k,t \rangle$-staleness.

Knowing whether a response is stale at read time requires strong consistency. Intuitively, by checking all possible values in the domain against a hypothetical staleness detector, we could determine the (strongly) consistent value to return. While we cannot do so synchronously, we *can* determine staleness asynchronously. Asynchronous staleness detection allows speculative execution [63] if a program contains appropriate compensation logic.

We first consider a staleness detector providing false positives. Recall that, in a Dynamo-style system, we wait for $R$ of $N$ replies before returning a value. The remaining $N-R$ replicas will still reply to the read coordinator. Instead of dropping these messages, the coordinator can compare them to the version it returned. If there is a mismatch, then either the coordinator returned stale data, there are in-flight writes in the system, or additional versions committed after the read. The latter two cases, relating to data committed after the response initiation, lead to false positives. In these cases, the read did not return "stale" data even though there were newer but uncommitted versions in the system. Notifying clients about newer but uncommitted versions of a data item is not necessarily bad but may be unnecessary and violates our staleness semantics. This detector does not require modifications to the Dynamo protocol and is similar to the read-repair process.

To eliminate these uncommitted-but-newer false positives (cases two and three), we need to determine the total, system-wide commit ordering of writes. Recall that replicas are unaware of the commit time for each version. The timestamps stored by replicas are not updated after commit, and commits occur after $W$ replicas respond. Thankfully, establishing a total ordering among distributed agents is a well-known problem that a Dynamo-style system can solve by using a centralized service [36] or using distributed consensus [43]. This requires modifications but is feasible.

## 5. EVALUATING DYNAMO $T$-VISIBILITY

As discussed in Section 3.4, PBS $t$-visibility depends on the propagation of reads and writes throughout a system. We introduced the *WARS* model as a means of reasoning about inconsistency in Dynamo-style quorum systems, but quantitative metrics such as staleness observed in practice depend on each of *WARS*'s latency distributions. In this section, we perform an analysis of Dynamo-style $t$-visibility to better understand how frequently "eventually consistent" means "consistent" and, more importantly, why.

PBS $k$-staleness is easily captured in closed form (Section 3.1). It does not depend on write latency or any environmental variables. Indeed, in practice, without expanding quorums or anti-entropy, we observe that our derived equations hold true experimentally.

$t$-visibility depends on anti-entropy, which is more complicated. In this section, we focus on deriving experimental expectations for PBS $t$-visibility. While we could improve the staleness results by considering additional anti-entropy processes (Section 4.2), we make the bare minimum of assumptions required by the *WARS* model. Conservative analysis decreases the number of experimental variables (supported by empirical observations from practitioners) and increases the applicability of our results.

### 5.1 Monte Carlo Simulation

In light of the complicated analytical formulation discussed in Section 4.1, we implemented *WARS* in an event-driven simulator for use in Monte Carlo methods. Calculating $t$-visibility for a given value of $t$ is straightforward. Denoting the $i$th sample drawn from distribution D as $D[i]$: draw $N$ samples from W, A, R, and S at time $t$, compute $w_t$, the $W$th smallest value of $\{W[i] + A[i], i \in [0, N)\}$, and check whether the first $R$ samples of R, ordered by $R[i] + S[i]$ obey $w_t + R[i] + t \leq W[i]$. This requires only a few lines of code. Extending this formulation to analyze $\langle k, t \rangle$-staleness given a distribution of write arrival times requires accounting for multiple writes across time but is not difficult.

### 5.2 Experimental Validation

To validate *WARS*, our simulator, and our subsequent analyses, we compared our predicted $t$-visibility and latency with measured values observed in a commercially available, open source Dynamo-style data store. We modified Cassandra to profile *WARS* latencies, disabled read repair (as it is external to *WARS*), and, for reads, only considered the first $R$ responses (often, more than $R$ messages would arrive by the processing stage, decreasing staleness). We ran

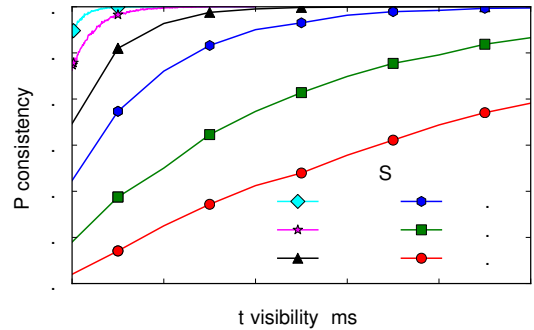

**Figure 4:** $t$-visibility with exponential latency distributions for W and A=R=S. Mean latency is $1/\lambda$. $N$=3, $R$=$W$=1.

Cassandra on three servers with 2.2GHz AMD Opteron 2214 dual-core SMT processors and 4GB of 667MHz DDR2 memory, serving in-memory data. To measure staleness, we inserted increasing versions of a key while concurrently issuing read requests.

Our *WARS* predictions matched our empirical observations of Cassandra's behavior. We injected each combination of exponentially distributed W $= \lambda \in \{0.05, 0.1, 0.2\}$ (means 20ms, 10ms and 5ms) and A=R=S $= \lambda \in \{0.1, 0.2, 0.5\}$ (means 10ms, 5ms and 2ms) across 50,000 writes. After empirically measuring the *WARS* distributions, consistency, and latency for each partial quorum configuration, we predicted the $t$-visibility and latency. Our average $t$-visibility prediction RMSE was $0.28\%$ (std. dev. $0.05\%$, max. $0.53\%$) for each $t \in \{1, \ldots, 199\}$ ms. Our predicted latency (for each of the $\{1.0, \ldots, 99.9\text{th}\}$ percentiles for each configuration) had an average N-RMSE of $0.48\%$ (std. dev. $0.18\%$, max. $0.90\%$). This validates our Monte Carlo simulator.

### 5.3 Write Latency Distribution Effects

As discussed in Section 4.1, the *WARS* model of Dynamo-style systems dictates that high one-way write variance (W) increases staleness. To quantify these effects, we swept a range of exponentially distributed write distributions (changing parameter $\lambda$, which dictates the mean and tail of the distribution) while fixing A=R=S.

Our results, shown in Figure 4, confirm this relationship. When the variance of W is $0.0625$ms ($\lambda = 4$, mean .25ms, one-fourth the mean of A=R=S), we observe a $94\%$ chance of consistency immediately after the write and $99.9\%$ chance after 1ms. However, when the variance of W is 100ms ($\lambda = .1$, mean 10ms, ten times the mean of A=R=S), we observe a $41\%$ chance of consistency immediately after write and a $99.9\%$ chance of consistency only after 65ms. As the variance and mean increase, so does the probability of inconsistency. Under distributions with fixed means and variable variances (uniform, normal), we observe that the mean of W is less important than its variance if W is strictly greater than A=R=S.

Decreasing the mean and variance of W improves the probability of consistent reads. This means that, as we will see, techniques that lower one-way write latency result in lower $t$-visibility. Instead of increasing read and write quorum sizes, operators could chose to lower (relative) W latencies through hardware configuration or by delaying reads. This latter option is potentially detrimental to performance for read-dominated workloads and may introduce undesirable queuing effects.

### 5.4 Production Latency Distributions

To study *WARS* in greater detail, we obtained production latency statistics from two internet-scale companies.

| %ile | Latency (ms) |
|---|---|
| 15,000 RPM SAS Disk | |
| Average | 4.85 |
| 95 | 15 |
| 99 | 25 |
| Commodity SSD | |
| Average | 0.58 |
| 95 | 1 |
| 99 | 2 |

**Table 1: LinkedIn Voldemort single-node production latencies.**

| %ile | Read Latency (ms) | Write Latency (ms) |
|---|---|---|
| Min | 1.55 | 1.68 |
| 50 | 3.75 | 5.73 |
| 75 | 4.17 | 6.50 |
| 95 | 5.2 | 8.48 |
| 98 | 6.045 | 10.36 |
| 99 | 6.59 | 131.73 |
| 99.9 | 32.89 | 435.83 |
| Max | 2979.85 | 4465.28 |
| Mean | 9.23 | 8.62 |
| Std. Dev. | 83.93 | 26.10 |
| Mean Rate | 718.18 gets/s | 45.65 puts/s |

**Table 2: Yammer Riak $N$=3, $R$=2, $W$=2 production latencies.**

| | |
|---|---|
| LNKD-SSD | $W = A = R = S$: <br> 91.22%: Pareto, $x_m = .235, \alpha = 10$ <br> 8.78%: Exponential, $\lambda = 1.66$ <br> N-RMSE: .55% |
| LNKD-DISK | $W$: <br> 38%: Pareto, $x_m = 1.05, \alpha = 1.51$ <br> 62%: Exponential, $\lambda = .183$ <br> N-RMSE: .26% |
| | $A = R = S$: LNKD-SSD |
| YMMR | $W$: <br> 93.9%: Pareto, $x_m = 3, \alpha = 3.35$ <br> 6.1%: Exponential, $\lambda = .0028$ <br> N-RMSE: 1.84% |
| | $A = R = S$: <br> 98.2%: Pareto, $x_m = 1.5, \alpha = 3.8$ <br> 1.8%: Exponential, $\lambda = .0217$ <br> N-RMSE: .06% |

**Table 3: Distribution fits for production latency distributions from LinkedIn (LNKD-*) and Yammer (YMMR).**

LinkedIn[3] is an online professional social network with over 135 million members as of November 2011. To provide highly available, low latency data storage, engineers at LinkedIn built Voldemort. Alex Feinberg, a lead engineer on Voldemort, graciously provided us with latency distributions for a single node under peak traffic for a user-facing service at LinkedIn, representing 60% read and 40% read-modify-write traffic [23] (Table 1). Feinberg reports that, using spinning disks, Voldemort is "largely IO bound and latency is largely determined by the kind of disks we're using, [the] data to memory ratio and request distribution." With solid-state drives (SSDs), Voldemort is "CPU and/or network bound (depending on value size)." As an aside, Feinberg also noted that "maximum latency is generally determined by [garbage collection] activity (rare, but happens occasionally) and is within hundreds of milliseconds."

Yammer[4] provides private social networking to over 100,000 companies as of December 2011 and uses Basho's Riak for some client data [3]. Coda Hale, an infrastructure architect, and Ryan Kennedy, also of Yammer, previously presented in-depth performance and configuration details for their Riak deployment in March 2011 [31]. Hale provided us with more detailed performance statistics for their application [30] (Table 2). Hale mentioned that "reads and writes have radically different expected latencies, especially for Riak." Riak delays writes "until the fsync returns, so while reads are often < 1ms, writes rarely are." Also, although we do not model this explicitly, Hale also noted that the size of values is important, claiming "a big performance improvement by adding LZF compression to values."

## 5.5 Latency Model Fitting

While the provided production latency distributions are invaluable, they are under-specified for *WARS*. First, the data are summary statistics, but *WARS* requires distributions. More importantly, the provided latencies are round-trip times, while *WARS* requires the constituent one-way latencies for both reads and writes. As our validation demonstrated, these latency distributions are easily collected, but, because they are not currently collected in production, we must fill in the gaps. Accordingly, to fit W, A, R, and S for each configuration, we made a series of assumptions. Without additional data on the latency required to read multiple replicas, we assume that each latency distribution is independently, identically distributed (IID). We fit each configuration using a mixture model with two distributions, one for the body and the other for the tail.

LinkedIn provided two latency distributions, whose fits we denote LNKD-SSD and LNKD-DISK for the SSD and spinning disk data. As previously discussed, when running on SSDs, Voldemort is network and CPU bound. Accordingly, for LNKD-SSD, we assumed that read and write operations took equivalent amounts of time and, to allocate the remaining time, we focused on the network-bound case and assumed that one-way messages were symmetric (W=A=R=S). Feinberg reported that Voldemort performs at least one read before every write (average of 1 seek, between 1-3 seeks), and writes to the BerkeleyDB Java Edition backend flush to durable storage either every 30 seconds or 20 MB—whichever comes first [23]. Accordingly, for LNKD-DISK, we used the same A=R=S as LNKD-SSD but fit W separately.

Yammer provided distributions for a single configuration, denoted YMMR, but separated read and write latencies. Under our IID assumptions, we fit single-node latency distributions to the provided data, again assuming symmetric A, R, and S. The data again fit a Pareto distribution with a long exponential tail. At the 98th percentile, the write distribution takes a sharp turn. Fitting the data closely resulted in a long tail, with 99.99+th percentile writes requiring tens of seconds—much higher than Yammer specified. Accordingly, we fit the 98th percentile knee conservatively; without the 98th percentile, the write fit N-RMSE is .104%.

We also considered a wide-area network replication scenario, denoted WAN. Reads and writes originate in a random datacenter, and, accordingly, one replica command completes quickly and the coordinator routes the others remotely. We delay remote messages by 75ms and apply LNKD-DISK delays once the command reaches a remote data center, reflecting multi-continent WAN delay [19].

We show the parameters for each distribution in Table 3 and plot each fitted distribution in Figure 5. Note that for $R$, $W$ of one, LNKD-DISK is not equivalent to WAN. In LNKD-DISK, we only have to wait for one of $N$ local reads (writes) to return, whereas, in WAN, there is only one local read (write) and the network delays all other read (write) requests by at least 150ms.

---

[3]LinkedIn. www.linkedin.com
[4]Yammer. www.yammer.com

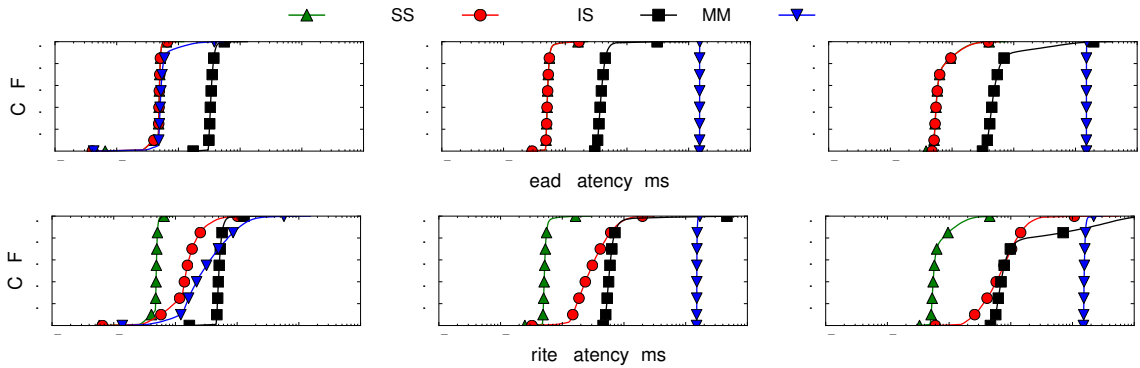

Figure 5: Read and write operation latency for production fits for $N{=}3$. For reads, LNKD-SSD is equivalent to LNKD-DISK.

### 5.6 Observed $t$-visibility

We measured the $t$-visibility for each distribution (Figure 6). As we observed under synthetic distributions in Section 5.3, the $t$-visibility depended on both the relative mean and variance of W.

LNKD-SSD and LNKD-DISK demonstrate the importance of write latency in practice. Immediately after write commit, LNKD-SSD had a 97.4% probability of consistent reads, reaching over a 99.999% probability of consistent reads after five milliseconds. LNKD-SSD's reads briefly raced its writes immediately after commit. However, within a few milliseconds after the write, the chance of a read arriving before the last write was nearly eliminated. The distribution's read and write operation latencies were small (median .489ms), and writes completed quickly across all replicas due to the distribution's short tail (99.9th percentile .657ms). In contrast, under LNKD-DISK, writes take much longer (median 1.50ms) and have a longer tail (99.9th percentile 10.47 ms). LNKD-DISK's $t$-visibility reflects this difference: immediately after write commit, LNKD-DISK had only a 43.9% probability of consistent reads and, ten ms later, only a 92.5% probability. This suggests that SSDs may greatly improve consistency due to reduced write variance.

We experienced similar effects with the other distributions. Immediately after commit, YMMR had a 89.3% chance of consistency. However, YMMR's long tail hampered its $t$-visibility increase and reached a 99.9% probability of consistency 1364 ms after commit. As expected, WAN observed poor chances of consistency until after the 75 milliseconds passed (33% chance immediately after commit); the client had to wait longer to observe the most recent write unless it originated from the reading client's data.

### 5.7 Quorum Sizing

In addition to $N{=}3$, we consider how varying the number of replicas (N) affects $t$-visibility while maintaining $R{=}W{=}1$. The results, depicted in Figure 7, show that the probability of consistency immediately after write commit decreases as $N$ increases. With 2 replicas, LNKD-DISK has a 57.5% probability of consistent reads immediately after commit but only a 21.1% probability with 10 replicas. However, at high probabilities of consistency, the wait time required for increased replica sizes is close. For LNKD-DISK, the $t$-visibility at 99.9% probability of consistency ranges from 45.3ms for 2 replicas to 53.7ms for 10 replicas.

These results imply that maintaining a large number of replicas for availability or better performance, results in a potentially large impact on consistency immediately after writing. However, the $t$-visibility staleness will still converge quickly.

### 5.8 Latency vs. $t$-visibility

Choosing a value for $R$ and $W$ is a trade-off between operation latency and $t$-visibility. To measure the obtainable latency gains, we compared $t$-visibility required for a 99.9% probability of consistent reads to the 99.9th percentile read and write latencies.

Partial quorums often exhibit favorable latency-consistency trade-offs (Table 4). For YMMR, $R{=}W{=}1$ results in low latency reads and writes (16.4ms) but high $t$-visibility (1364ms). However, setting $R{=}2$ and $W{=}1$ reduces $t$-visibility to 202ms and the combined read and write latencies are 81.1% (186.7ms) lower than the fastest strict quorum ($W{=}1$, $R{=}3$). A 99.9% consistent $t$-visibility of 13.6ms reduces LNKD-DISK read and write latencies by 16.5% (2.48ms). For LNKD-SSD, across $10M$ writes ("seven nines"), we did not observe staleness with $R{=}2$, $W{=}1$. $R{=}W{=}1$ reduced latency by 59.5% (1.94ms) with a corresponding $t$-visibility of 1.85ms. Under WAN, $R > 1$ or $W > 1$ results in a large latency increase because this requires WAN messages. In summary, lowering values of $R$ and $W$ can greatly improve operation latency and that $t$-visibility can be low even when we require a high probability of consistent reads.

## 6. DISCUSSION AND FUTURE WORK

In this section, we discuss enhancements to partial quorum systems that PBS enables along with future work for PBS.

**Latency/Staleness SLAs.** With PBS, we can automatically configure replication parameters by optimizing operation latency given constraints on staleness and minimum durability. Data store operators can subsequently provide service level agreements to applications and quantitatively describe latency/staleness trade-offs to users. Operators can dynamically configure replication using on-line latency measurements. PBS provides a quantitative lens for analyzing consistency guarantees that were previously unknown. This optimization formulation is likely non-convex, but the state space for configurations is small ($O(N^2)$). This optimization also allows disentanglement of replication for reasons of durability from replication for reasons of low latency and higher capacity. For example, operators can specify a minimum replication factor for durability and availability but can also automatically increase $N$, decreasing tail latency for fixed $R$ and $W$.

**Variable configurations.** We have assumed the use of a single replica configuration ($N$, $R$, and $W$) across all operations. However, one could vary these operations over time and across keys. By specifying a target latency, one could periodically modify $R$ and $W$ to more efficiently guarantee a desired bound on staleness, or vice versa. These time-varying configurations require additional refinements and revisit prior work on fluid replication [53].

**Stronger guarantees.** We have focused on bounded staleness analysis, but there are other (often stronger) forms of consistency (such as causal consistency) [61]. Predicting the probability of attaining more complex consistency semantics requires additional modeling of application access patterns. This is possible, but we suspect that modeling the *worst-case* semantics of these operations will result in unfavorably low probabilities of consistent operations. We can see this in Aiyer et al.'s analysis of Byzantine $k$-quorums [11]: in a worst-case deployment, with an adversarial scheduler, the lower bound on guaranteed recency is high. We conjecture that the bound would be even higher had the authors performed an analysis of stronger consistency models.

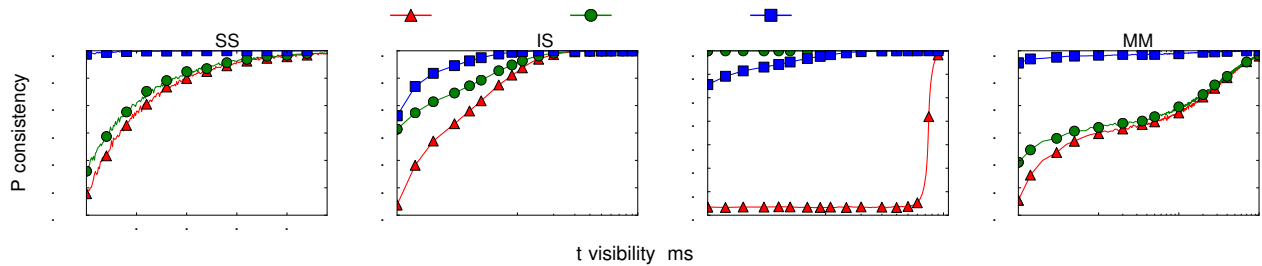

Figure 6: $t$-visibility for production operation latencies.

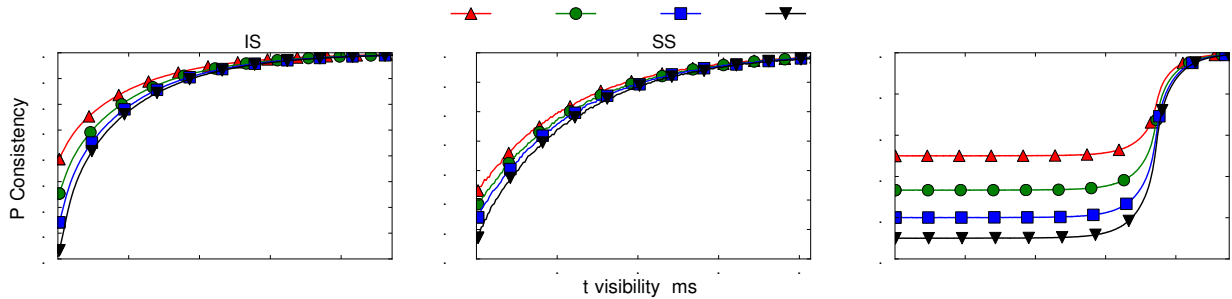

Figure 7: $t$-visibility for production operating latencies for variable $N$ and $R=W=1$.

**Alternative architectures.** Dynamo is conceptually easy to understand and implement (*WARS*) but is painful to analytically analyze. Is there a design that finds a better middle ground between operational elegance and simplicity of analysis within the eventually consistent design space? Prior work on deterministic bounded staleness (Section 7) provides guidance but often sacrifices availability and may be more complex to reason about.

**Multi-key operations.** We have considered single-key operations, however the ability to perform multi-key operations is potentially attractive. For read-only transactions, if the key distribution is random and each quorum is independent, we can multiply the staleness probabilities of each key to determine multi-key staleness probabilities. Achieving atomicity of writes to multiple keys requires more complicated coordination mechanisms such as two-phase commit, increasing operation latency. Transactions are feasible but require considerable care in implementation, complicating what is otherwise a simple replication scheme.

**Failure modes.** In our evaluation of $t$-visibility, we focused on normal, steady state operating conditions. Unless failures are common-case, they affect tail staleness probabilities (which appear as latency spikes in *WARS*). For example, if, as Jeff Dean of Google suggests [19], servers crash at least twice per year, given a ten-hour downtime per failure, this results in .23% downtime per machine per year. If failures are correlated, this may be a problem. If they are independent, a replica set of $N$ nodes with $F$ failed nodes behaves like an $N - F$ replica set. The probability of all $N$ nodes failing is $(.23)^N\%$ ("five nines" reliability for $N=3$) and the probability tail will hide these failures. Quantifying these effects requires information about failure rates and their impact on latency distributions but would be beneficial. Modeling recovery semantics such as hinted handoff [20, Section 4.6] would also be useful.

## 7. RELATED WORK

We surveyed quorum replication techniques [7, 9, 8, 10, 11, 26, 29, 34, 37, 49, 51, 52] in Section 2. In this work, we specifically draw inspiration from probabilistic quorums [49] and deterministic $k$-quorums [10, 11] in analyzing expanding quorum systems and their consistency. We believe that revisiting probabilistic quorum systems—including non-majority quorum systems such as tree quorums—in the context of write propagation, anti-entropy, and Dynamo is a promising area for theoretical work.

Data consistency is a long-studied problem in distributed systems [18] and concurrent programming [35]. Given the CAP Theorem and the inability to maintain all three of consistency, availability, and partition tolerance [27], data stores have turned to "eventually consistent" semantics to provide availability in the face of partitions [18, 61]. Real-time causal consistency is the strongest consistency model achievable in an available, one-way convergent (eventually consistent) system [48]. However, there is a plethora of alternative consistency models offering different performance trade-offs, from session guarantees [60] to causal+ consistency [46] and parallel snapshot isolation [57]. Instead of proposing a new consistency model and building a system implementing new semantics, we have examined what consistency existing, widely deployed quorum-replicated systems actually provide.

Prior research examined how to provide deterministic staleness bounds. FRACS [67] allows replicas to buffer updates up to a given staleness threshold under multiple replication schemes, including master-drive and group gossip. AQuA [40] asynchronously propagates updates from a designated master to replicas that in turn serve reads with bounded staleness. AQuA actively selects which replicas to contact depending on response time predictions and a guaranteed staleness bound. TRAPP [54] provides trade-offs between precision and performance for continuously evolving numerical data. TACT [65, 66] models consistency along three axes: numerical error, order error, and staleness. TACT bounds staleness by ensuring that each replica (transitively) contacts all other replicas in the system within a given time window. Finally, PIQL [13] bounds the number of operations performed per query, trading operation latency at scale with the amount of data a particular query can access, impacting accuracy. These deterministically bounded staleness systems represent the deterministic dual of PBS.

|  | LNKD-SSD | | | LNKD-DISK | | | YMMR | | | WAN | | |
|---|---|---|---|---|---|---|---|---|---|---|---|---|
|  | $L_r$ | $L_w$ | $t$ | $L_r$ | $L_w$ | $t$ | $L_r$ | $L_w$ | $t$ | $L_r$ | $L_w$ | $t$ |
| $R{=}1, W{=}1$ | **0.66** | **0.66** | **1.85** | 0.66 | 10.99 | 45.5 | 5.58 | 10.83 | 1364.0 | **3.4** | **55.12** | **113.0** |
| $R{=}1, W{=}2$ | 0.66 | 1.63 | 1.79 | 0.65 | 20.97 | 43.3 | 5.61 | 427.12 | 1352.0 | 3.4 | 167.64 | 0 |
| $R{=}2, W{=}1$ | **1.63** | **0.65** | **0** | **1.63** | **10.9** | **13.6** | **32.6** | **10.73** | **202.0** | 151.3 | 56.36 | 30.2 |
| $R{=}2, W{=}2$ | **1.62** | **1.64** | **0** | 1.64 | 20.96 | 0 | 33.18 | 428.11 | 0 | 151.31 | 167.72 | 0 |
| $R{=}3, W{=}1$ | 4.14 | 0.65 | 0 | **4.12** | **10.89** | **0** | **219.27** | **10.79** | **0** | **153.86** | **55.19** | **0** |
| $R{=}1, W{=}3$ | 0.65 | 4.09 | 0 | 0.65 | 112.65 | 0 | 5.63 | 1870.86 | 0 | 3.44 | 241.55 | 0 |

Table 4: $t$-visibility for $p_{st} = .001$ (99.9% probability of consistency for $50,000$ reads and writes) and 99.9th percentile read ($L_r$) and write latencies ($L_w$) across $R$ and $W$, $N{=}3$ ($1M$ reads and writes). Significant latency-staleness trade-offs are in bold.

Finally, recent research has focused on measuring and verifying the consistency of eventually consistent systems both theoretically [28] and experimentally [16, 62]. This is useful for validating consistency predictions and understanding staleness violations.

## 8. CONCLUSION

In this paper, we introduced Probabilistically Bounded Staleness, which models the expected staleness of data returned by eventually consistent quorum-replicated data stores. PBS offers an alternative to the all-or-nothing consistency guarantees of today's systems by offering SLA-style consistency predictions. By extending prior theory on probabilistic quorum systems, we derived an analytical solution for the $k$-staleness of a partial quorum system, representing the expected staleness of a read operation in terms of versions. We also analyzed $t$-visibility, or expected staleness of a read in terms of real time, under Dynamo-style quorum replication. To do so, we developed the *WARS* latency model to explain how message reordering leads to staleness under Dynamo. To examine the effect of latency on $t$-staleness in practice, we used real-world traces from internet companies to drive a Monte Carlo analysis. We find that eventually consistent quorum configurations are often consistent after tens of milliseconds due in large part to the resilience of Dynamo-style protocols. We conclude that "eventually consistent" partial quorum replication schemes frequently deliver consistent data while offering significant latency benefits.

## Interactive Demonstration

An interactive demonstration of Dynamo-style PBS is available at http://pbs.cs.berkeley.edu/#demo.

## Acknowledgments

The authors would like to thank Alex Feinberg and Coda Hale for their cooperation in providing real-world distributions for experiments and for exemplifying positive industrial-academic relations through their conduct and feedback.

The authors would also like to thank the following individuals whose discussions and feedback improved this work: Marcos Aguilera, Peter Alvaro, Eric Brewer, Neil Conway, Greg Durrett, Jonathan Ellis, Andy Gross, Hariyadi Gunawi, Sam Madden, Bill Marczak, Kay Ousterhout, Vern Paxson, Mark Phillips, Christopher Ré, Justin Sheehy, Scott Shenker, Sriram Srinivasan, Doug Terry, Greg Valiant, and Patrick Wendell. We would especially like to thank Bryan Kate for his extensive comments and Ali Ghodsi, who, in addition to providing feedback, originally piqued our interest in theoretical quorum systems.

This work was supported by gifts from Google, SAP, Amazon Web Services, Blue Goji, Cloudera, Ericsson, General Electric, Hewlett Packard, Huawei, IBM, Intel, MarkLogic, Microsoft, NEC Labs, NetApp, NTT Multimedia Communications Laboratories, Oracle, Quanta, Splunk, and VMware. This material is based upon work supported by the National Science Foundation Graduate Research Fellowship under Grant DGE 1106400, National Science Foundation Grants IIS-0713661, CNS-0722077 and IIS-0803690, the Air Force Office of Scientific Research Grant FA95500810352, and by DARPA contract FA865011C7136.